\documentclass[%
 reprint,
 amsmath,amssymb,
 aps,
]{revtex4-2}

\usepackage{graphicx}
\usepackage{caption}
\usepackage{dcolumn}
\usepackage{bm}

\begin{document}

\preprint{APS/123-QED}


\title{Including stress relaxation in point-process model for seismic occurrence}

\author{Giuseppe Petrillo}
\affiliation{
  The Institute of Statistical Mathematics, Research Organization of Information and Systems, Tokyo, Japan
}%

\author{Eugenio Lippiello}
\affiliation{
  Department of Mathematics and Physics, University of Campania ``L. Vanvitelli'', Viale  Lincoln 5, 81100 Caserta, Italy
}%
\author{Jiancang Zhuang}
\affiliation{
  The Institute of Statistical Mathematics, Research Organization of Information and Systems, Tokyo, Japan
}%

\date{\today}

\begin{abstract}
Physics-based and statistic-based models for describing seismic occurrence are two sides of the same coin. In this article we compare the temporal organization of events obtained in a spring-block model for the seismic fault with the one predicted by probabilistic models for seismic occurrence. Thanks to the optimization of the parameters, by means of a Maximum Likelihood Estimation, it is possible to identify  the statistical model which fits better the physical one. The results show that the best statistical model must take into account the non trivial interplay between temporal clustering, related to aftershock occurrence, and the stress discharge following the occurrence of high magnitude mainshocks. The two mechanisms contribute in different ways according to the minimum magnitude considered in the data fitting catalog. 
\end{abstract}

\maketitle


\section{\label{sec:level1} INTRODUCTION}

The description and understanding of seismic phenomena is fascinating not only to add a piece to basic physics knowledge, but also for the social impact it can have. Seismic prediction is a longstanding and debated topic in the geophysical community, with different scenarios, ranging from the extreme position of a completely random unpredictable (Poisson) process, up to deterministic predictions \cite{main99}. The difficulty in answering this debate can be attributed to the complexity of the earthquake process, presenting different mechanisms acting on very different spatial and temporal scales \cite{dAGGL16}. \\
Currently, epidemic-like probabilistic models \cite{ogata88,ogata1998,PZ22} have given and continue to provide a fundamental contribution to seismic forecasting. These class of models are able to capture the spatio-temporal clustering of earthquakes, with the majority of earthquakes, usually termed aftershocks, occurring soon after large earthquakes and close in space to their epicenter. Within these models seismic hazard increases after the occurrence of large earthquakes. Nevertheless it is not possible to achieve information on the timing of the next large earthquake in a given region because magnitudes are randomly assigned to each earthquake from the empirical magnitude distribution and does not depend on the previous seismic history. 
On the other hand, physical models originally implement the idea that an earthquake is the sudden relaxation of the stress accumulated over years or decades in such a region. Accordingly one could expect that the region experiencing the largest slip also relaxes most of the accumulated stress and therefore is less prone to host a new earthquake, as in the original hypothesis of alternation proposed by Gilbert in 1894 \cite{Gil09}. Subsequent models, developed along this direction, assume that consecutive earthquakes substantially re-rupture the same fault segment, nucleating characteristic earthquakes which are roughly equal in size and roughly periodic is time. These models usually termed seismic gap or seismic cycle models \cite{MNSK79,Nis91}, propose the possibility of quasi deterministic prediction of the timing of the next large earthquake. 
Testing of the seismic gap hypothesis have shown its inefficiency \cite{KJ91,RJK03} even if it is still sometimes used in earthquake forecasting \cite{MSG17}. Furthermore the seismic gap model does not provide any justification for the occurrence of aftershocks which conversely, as anticipated, is a very distinctive feature of seismic catalogs.  
Nevertheless, aftershock occurrence, and as a consequence also short-term clustering,  is not incompatible within the seismic gap hypothesis if aftershocks are located outside the region involved during the mainshock slip or, at most, in regions with low levels of the mainshock slip. This feature has been documented for $101$ large subduction zone plate boundary mainshocks with well determined coseismic slip distributions \cite{WTBK18}.
This study has indeed revealed a deficit of aftershocks  inside the mainshock slip area, consistently with the hypothesis of large slip areas re-locking. As a further indication of compatibility between the seismic gap hypothesis and aftershock occurrence is the observation that larger aftershocks occur farther away than smaller aftershocks \cite{ES15}. \\
Interestingly, the interplay between the stress accumulation/relaxation mechanism underlying the seismic gap hypothesis and the occurrence of aftershocks described by epidemic models, has been enlightened in a spring-block model for the seismic fault.  The model, is a generalization of the OFC model \cite{OFC}, which represents a cellular automaton version of spring-block model for seismic fault \cite{BK67}. In particular, the considered model, defined afterslip relaxation OFC (AROFC) model, belongs to the class of models \cite{HZK99,jagla2010,JK10,JLR14,ZS16,BD18,LPLR19,PLLR20,LPLR21} which add a relaxation mechanism introducing aftershock occurrence within the OFC model \cite{OFC}. In particular the AROFC model describes the seismogenic fault as an elastic layer, like in the OFC description, and it takes into account its coupling with  a more ductile layer where stress is gradually relaxed by a post-seismic deformation, usually termed afterslip \cite{PA04,PA07}. The model is very efficient in quantitatively reproducing the main statistical features of seismic catalogs, as well as to explain the  connection between aftershocks and afterslip \cite{LPLR19,LPLR21} or the effects of Moho depth on magnitude distribution \cite{GTPBL22}. 
Interestingly, in the AROFC model \cite{PRL22}, even if the higher seismic rate is observed after the occurrence of large mainshock, some features of the seismic gap hypothesis are recovered, with large  earthquakes usually occurring in gap regions. Nevertheless, the timing of mainshocks is irregular and weakly depends on the time delay of that region from the previous slip. Recent results therefore show that epidemic description and gap hypothesis, in a weaker form, can coexist.\\
The main purpose of this study is to combine the two mechanisms in a statistical model for seismic occurrence which, on one side,  preserves the epidemic description and, on the other side, keeps into account the stress accumulation/relaxation mechanism beyond the seismic gap hypothesis. More precisely we show how the two scenarios can be combined in various statistical models \cite{vere88,vere,lomnitz1966,SB00}. The problem is that an accurate statistical validation of the different models is not possible in instrumental seismic catalogs since the occurrence of large earthquakes, with overlapping slip regions, is a rare event. 
For this reason we search for the statistical model which better reproduce the temporal organization of earthquakes in the AORFC model, which conversely allows us to generate arbitrarily long simulated catalogs for a full statistical validation. More precisely we identify the best parameters for each model via a Maximum Likelihood Estimation (MLE) finding a non-trivial dependence of the Log-Likelihood function on the minimum magnitude of the catalog.  \\
The paper is organized as follows. Section \ref{sec1} is an overview of the AROFC model that is used for the generation of the synthetic catalog. The section \ref{sec2} is dedicated to the introduction of the various statistical models that are used subsequently. The optimization of the parameters by means of MLE and the comparison of the efficiency of the different models in fitting the data, are carried out in the following sections. The last section is devoted to conclusions.

\section{THE AROFC MODEL}
\label{sec1}

    The model is composed by two layers: a layer $H$ and a layer $U$, which mimics the brittle and the ductile part of the fault, respectively. The layer $U$ is driven by the tectonic dynamics at the velocity $V_D$ and is elastically coupled to a layer $H$. Each layer is an object that can be considered as a set of blocks organized on a lattice of size $L_x \times L_y$. For simplicity, we assume $V_D$ directed along the $x$-direction and, therefore, the displacement is confined along $x$, defined $h_i(t)$ in the layer $H$ and $u_i(t)$ in the layer $U$. We assume a short-range interaction where each block is elastically connected only with its nearest neighbor blocks. Within this approximation, the total stress on block $H$ at position $i$ can be divided into two contributions: the intra-layer stress $f_i=k_h \Delta h_i$ and the inter-layer stress $g_i = k(u_i-h_i)$. In the last expressions, $\Delta $ is the discretized Laplacian, whereas $k_h$ and $k$ represents the elastic constants of the inter-layer and the intra-layer interactions, respectively. Regarding the friction force, we assume two different forms. For the brittle fault $H$ we adopt a Coulomb failure criterion: as soon as the total force $f_i + g_i$ overcomes a local frictional threshold $f^{th}_i$, the position $h_i$ becomes unstable and the $i-$th block performs a slip of fix quantity $\Delta h$. The stress evolves as follow: 
    \begin{eqnarray}
        &f_i \rightarrow f_i - 4 k_h\Delta h \nonumber \\
        &f_i \rightarrow f_j + k_h \Delta h  \nonumber \\
        &g_i \rightarrow g_i - 4k_h \zeta \Delta h  \nonumber \\
        &g_j \rightarrow g_j - k_h (\zeta-\epsilon) \Delta h,
    \end{eqnarray}
where $\zeta=(1-q_0)(\frac{k}{4k_h})$ and $\epsilon=(1-q_0-4q_1)(\frac{k}{4k_h})$. The first quantity takes into account the coupling between the two layers, while the second one is the dissipation. The quantity $k_h \Delta h = \Delta f$ represents the stress drop and it is extracted from a Gaussian distribution with average $\langle \Delta f \rangle$ and standard deviation $\sigma$. For the ductile layer, we assume a velocity-strengthening friction, taking the stationary form of the rate-and-state friction law $f_i^U (t) = \sigma_N (\mu_c + A \log(\frac{u_i(t)}{V_D}))$, where $\sigma_N$ is the normal stress, $\mu_c$ is the friction coefficient when the block $U$ slides at the steady velocity $V_D$ and $A>0$ for a velocity-strengthening material. The slip $\Delta h$ of $h_i$ induces a coseismic slip $q_0 \Delta h$ of $u_i$ and $q_1 \Delta h$ of the blocks $u_j$ (where $j$ is a nearest neighbor of $u_i$). When the total stress acting on $i$-th block of layer $H$ exceeds the local frictional stress threshold $f^{th}_i$, i.e., when $f_i +g_i \ge f^{th}_i$. The in-depth explanation of the dynamics of the model was studied in \cite{LPLR19,PLLR20,LPLR21}. In this article we simulate a fault with system size $L_y=200$ and $L_x=400$ with absorbing boundary conditions. \\
An earthquake is defined as a series of slips starting from an initial instability of the block $i$, whose position defines the epicentral coordinates of the earthquake and terminating when $f_i +g_i \le f^{th}_i$. Taking into account that the  $i$-th block can slip more than one time during an earthquake, we introduce the quantity $n_k(i)$ for the number of slips performed by the $i$-th block during the $k$-th earthquake. 
The seismic moment $M_k$ released during the $k$-th earthquake can be defined as  $M_k \propto \sum_i n_k(i) \Delta h$, where the sum extends over all lattices sites. We finally define the moment magnitude $m_k=(2/3) \log_{10}M_k$, where we have set to zero the arbitrary additive constant. \\
Therefore, with the model described, it is possible to obtain as output a collection of the magnitude of events as a function of time, called the seismic catalogue. We want to emphasize that this model reproduces all the statistical laws of earthquakes correctly, such as Gutenberg-Richter law, Omori-Utsu law, $m$ vs $Log(A)$- scaling, etc. \cite{PLLR20}. \\

\begin{figure}[b]
\includegraphics[width=8cm]{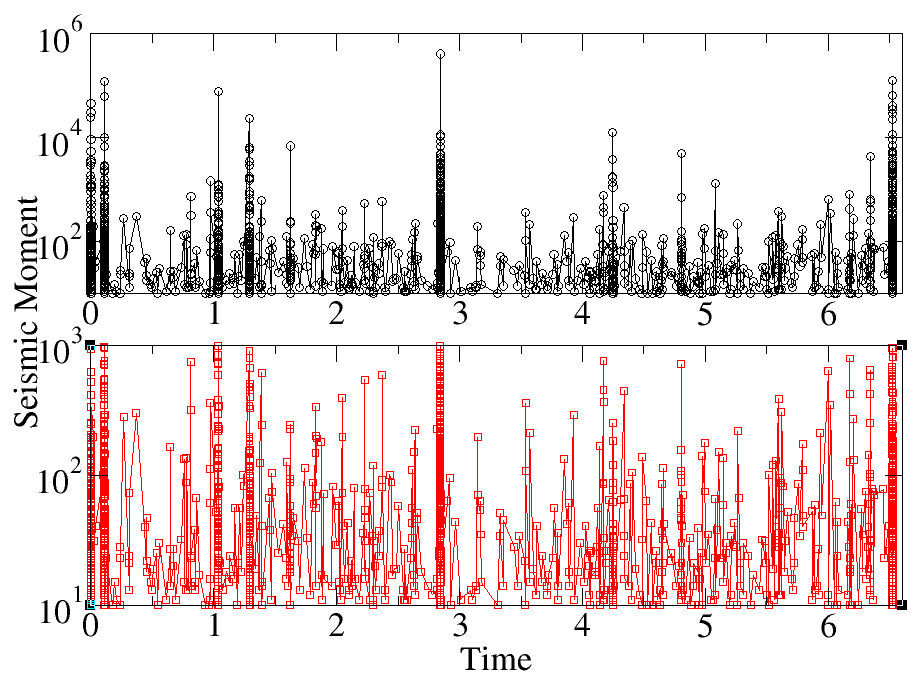}
\caption{\label{fig:epsart} The numerical earthquake catalogues obtained with the OFCR model. (Top panel) Black circles indicate the seismic moment $M_i$ of all the event $i$ occurred at time $t_i$. (Bottom panel) Red squares indicate the seismic moment $M_i$ of all the event $i$ occurred at time $t_i$ with the extra constraint $M_i \leq 1000$.}
\label{catalog}
\end{figure}
We simulated two synthetic catalogues: the first, $CAT_1$, consists of $N = 1E6$ events with no constraint on the seismic moment; the second, $CAT_2$, is a subset of $CAT_1$ obtained by considering only earthquakes with a seismic moment $M$ less than $1000$. In particular in Fig.(\ref{catalog}) the seismic moment versus the occurrence time is shown. Here, each symbol corresponds to an event with occurrence time $t$ and moment magnitude $M_k$. It is immediate to observe the presence of correlated sequences, called clusters, in the catalog. Each cluster is composed by a mainshock (defined as the biggest event of the sequence) and some aftershocks/foreshocks (all the events occurred after/before the mainshock). It is common to observe a cluster with the first event bigger than all the others, this case is not a pathology of the model, but represents a single seismic sequence without the occurrence of foreshocks, which is often observed in instrumental catalogs. \\

\section{SELF TRIGGERING AND STRAIN-RELEASE MODELS}
\label{sec2}
    A Point Process (PP) is a very popular mathematical model used in seismology, especially for earthquake forecasting, implementing the assumption that an earthquake is an isolated point in time. PP can be defined by means of the number of events $\mathcal{N}$ occurred in a certain time interval $\Delta t$. If this time span is small, $\mathcal{N}$ can assume a null value (event not happened) or a unit value (event happened). Therefore the process is univocally determined by the probability of observing an event in a certain interval $\Delta t$ conditioned to the previous history $\mathcal{H}_t$. The limit

    \begin{equation}
        \lambda(t|\mathcal{H}_t) = \lim_{\Delta t \rightarrow 0} \frac{P(\mathcal{N}_{\Delta t}=1|\mathcal{H}_t)}{\Delta t}
    \end{equation}
    then represents the probability density (usually called "intensity function") of observing an earthquake at a certain time $t$, given its history up to time $t$. \\
    Forecasting models also consider a point-process description for the spatial occurrence of earthquakes. In this study, for simplicity, we only consider temporal models.
    
    \subsection{Self-Correcting Models}
        In the classical rebound model, one assumes that stress slowly builds up to the breaking point when an earthquake occurs. The rupture of the fault following this event, on the other hand, resulted in a reduction in stress. In general, the conditional intensity rate of this model can be written as: 
        \begin{equation}
            \lambda_{SR}(t|\mathcal{H}_t) = e^{F(t) - G[0,t)} 
            \label{sc}
        \end{equation}
        where $G[0,t)$ is an history-related function and $F(t)$ is the stress loading which, in general, is considered a linear increasing function of the time.
        The exponential form is chosen to ensure positivity of the rate $ \lambda_{SR}$. \\  
        The Stress Release Model (SR) belongs to the self-correcting models introduced by \cite{vere}. In practice, the elastic strain is accumulated due to the long-term tectonic loading, and is released when it exceeds a certain threshold level. After the release of the stress, a certain period is needed to the re-accumulation of the stress and the genesis of a subsequent event. Therefore Eq.(\ref{sc}) for a SR model, as proposed by \cite{vere,VD88,LHB99, ZV91}, can be written as       
        \begin{equation}
            \lambda_{SR}(t|\mathcal{H}_t) = e^{\alpha + \beta t - \xi \mathcal{S}[0,t)} 
            \label{sr}
        \end{equation}
        where $F(t)=\alpha + \beta t$ is a linear function of the time, representing the tectonic loading, whilst $ \xi \mathcal{S}[0,t)$ is related to the reduction of stress due to the past history. Indicating by $M_i$ the seismic moment of the $i$-th event, $\mathcal{S}[0,t) = \sum_{i:t_i < t} S_i$, where $S_i$ is the stress released by the $i$-th earthquake, and the sum extends to all events occurred at $t_i < t $. A reasonable hypothesis \cite{BV93,LPT21} is that  $S_i$,is proportional to square root of the energy released $S_i \propto \sqrt{E_i}$, leading to $S_i \propto \sqrt{M_i}$. Therefore, the fitting parameter set for SR model in Eq.(\ref{sr}) is $\Theta_{SR} = (\alpha,\beta,\xi)$. \\
        Sometimes a simplified version of the SR model is used in the description of seismicity. In particular, in refs.\cite{IW79,SB00} is has assumed that, on average, the total stress released is proportional to the number of occurred earthquakes, leading to   $\mathcal{S}[0,t) = \mathcal{N}[0,t) $, where $\mathcal{N}[0,t)$ denotes the number of earthquakes in the interval $(0, t]$. With this position, we obtain   
        \begin{equation}
            \lambda_{SC}(t|\mathcal{H}_t) = e^{\alpha + \beta t - \xi \mathcal{N}[0,t)}
            \label{SC}
        \end{equation}
        The fitting parameter set is the same of the SR model and, from now on, we call this model simply Self-Correcting (SC) model since it does not take into account the energy released by previous earthquakes.
    
    \subsection{Triggering models}
    
    \textit{Self-exiting} point process (SEPP) have been also used to model the seismic occurrence and for seismic forecasting. A typical Hawkes process \cite{hawkes1973} can be written as
    
    
     \begin{equation}
         \lambda(t|\mathcal{H}_t) = \mu(t) + \sum_{j:t_j<t} \nu(t-t_j),
         \label{hawkes}
    \end{equation}
    where $\mu(t)$ represent the background seismicity and $\nu(t)$ the triggering intensity function, in practice, $\nu$ controls the clustering density.  
    For the functional form of $\nu(t)$, many proposals have been made \cite{EL66,ogata88,UORM95,VO82,OA82,lomnitz1966}. \\
    Eq.(\ref{hawkes}) describes seismicity as a branching process where each earthquake can have aftershocks, and these can have their own aftershocks, leading to a cascade seismic process. \\
    In this paper we focus on two types of triggering form. The first one, suggested by the elastic aftereffect theory, gives for the probability density $\lambda_{TR1}(t|\mathcal{H}_t)$
     \begin{equation}
         \lambda_{TR1}(t|\mathcal{H}_t) = e^\alpha + \sum_{j:t_j<t} \phi e^{-\theta(t-t_j)},
         \label{hawkes2}
    \end{equation}
     where we have placed $\mu(t)=e^\alpha$ and $\nu(t) = \phi e^{-\theta t}$. The second one is the ETAS model which implements the Omori-Utsu law $\nu(t) \propto \phi (t+c)^{-\theta} $
    \begin{equation}
         \lambda_{ETAS}(t|\mathcal{H}_t) = e^\alpha + \sum_{j:t_j<t} \frac{\phi}{(t-t_j+c)^\theta},
         \label{hawkes3}
    \end{equation}
    The parameters to fit in the triggering models depend on the functional choice of $\nu$.

    \subsection{The ETASLC model}
        We propose to combine triggering with self-correcting models to obtain the so-called \textit{Epidemic Type Aftershock Sequence Long-term Correcting} (ETAFLC) model. \\
        Analytically, two terms of the conditional intensity rate must be modified. The first step is to set the background seismicity $\mu(t)$ of the triggering models with the temporal stress loading proposed in Eq.(\ref{hawkes2}), that is
        
        \begin{equation}
            \mu(t) = e^{\alpha + \beta t}
        \end{equation}
        while, the stress discharge mechanism after a large earthquake can be reasonably implemented in the functional form of $\nu(t)$ simply subtracting a constant quantity $\xi$, which represents the new parameters to be fitted as in Eqs.(\ref{sr},\ref{SC}). \\
        Overall, putting all this information together, employing the exponential triggering form, the conditional rate of the ETAFLC model can be written as:      
        \begin{equation}
            \lambda_{ETAFLC1}(t|\mathcal{H}_t) = e^{\alpha +\beta t} + \sum_{j:t_j<t} \left( \phi e^{-\theta(t-t_j)} -\xi \right),
        \end{equation}
        while, implementing the Omori-Utsu law for the triggering part       
         \begin{equation}
            \lambda_{ETAFSLC2}(t|\mathcal{H}_t) = e^{\alpha +\beta t} + \sum_{j:t_j<t} \left( \frac{\phi}{(t-t_j+c)^\theta} -\xi \right).
        \end{equation}
        It is evident that the constant $\xi$, present inside the sum, is responsible for a correcting term proportional to the number of earthquakes occurred up to time $t$, playing a similar role of $\mathcal{N}[0,t)$ in Eq.(\ref{SC}).

\section{LOG-LIKELIHOOD (LL) MAXIMIZATION}

For asynchronous data, the likelihood for a non-stationary Poisson process, characterized by an intensity function $\lambda(t)$, is
\begin{equation}
    L(Y|\Theta) = \prod_{i=1}^N \lambda(t_i) e^{-\int_{t_{in}}^{t_{fin}} \lambda(t)dt}
\end{equation}
where $Y = (y_1,...,y_N)$ are the $N$ observed data, $t_i$ is the occurrence time of the $i$-th event in the time interval $[t_{in},t_{fin}]$ and $\Theta$ is the parameter set. \\
The optimization of the parameters, for both models, is carried out by a Maximum Likelihood Estimation (MLE) \cite{ogata1998,LGdAMG14,ogata1983,oz06}. The parameter optimisation procedure is carried out via a component-wise Markov-Chain-Monte-Carlo (MCMC) procedure. The MLE algorithm used is presented in App.\ref{app1}. \\
The model that best reproduces the data of the simulated catalog is the one with the minimum AIC value. The AIC value of a certain model is $AIC = 2k - 2LL$, where $k$ is the number of the parameters in the model. In practice, this criterion avoids under-fitting by choosing the model with the highest LL, but assigns a penalty for each additional parameter.

\section{COMPARING MODEL AND SIMULATIONS}

For the AROFC catalogue, the parameter estimates for the different fitting models are reported in Tabs.(\ref{param},\ref{param2}), the set up values for the MCMC algorithm are shown in Tab.(\ref{tabpar}) in the App.\ref{app2}. In Figs.(\ref{llsc},\ref{llsr},\ref{lltr1},\ref{lltr2},\ref{llselc1},\ref{llselc2}) we plot the evolution of parameters during the optimization parameter, for all the models considered. The algorithm is carried out with $10^6$ Monte Carlo steps and it is also very clear that the algorithm converges to a stationary value for all parameters and all models. 
\\
We first analyse the complete catalogue and results of Tab.(\ref{param}) clearly show that the triggering model is better than both the self-correcting model and the stress release model, but combining the two previous models results in an $AIC_{ETASLC2}$ much smaller than all the others, moreover, both ETAS and ETASLC2 have an AIC value smaller than corresponding model defined implementing an exponential triggering function. This results give us two important information. The first, as already well known, is that the Omori-Utsu form is the most plausible triggering function $\nu(t)$ for the description of the short term triggering effects. This also confirms the correct implementation in the physical model of the relaxation function after each event. The second, that both ETASLC1 to ETASLC2 models over-perform their respective triggering models.
Interestingly, if we carry out the same study considering only the events of the numerical catalog below a certain seismic moment $ M_{max}=1000$, we observe that the AIC values for both ETASLC models dramatically increase becoming comparable to the triggering model ETAS. This is strong evidence that events of larger magnitude contribute to a much greater decrease in stress than do small events alone, in fact, the AIC values of the triggering models remain almost unchanged (see Tab.(\ref{param2})) restricting the catalogue magnitude. Again, a comparison of the triggering models TR1 and ETAS shows how the Omori-Utsu functional form is a more appropriate choice for describing short-term seismicity. The results also confirm the recent findings of \cite{PRL22} which show a slight dependence of the recurrence times two subsequent mainshocks. 
The magnitude dependence is also observed by looking at the simple Self-Correcting and Stress Release models. In fact, only in the case in which large events are considered, we note an improvement of the AIC for the SR model, which contains precisely the information on the magnitude of the earthquake. In the case of minor events, however, the AIC remains almost unchanged between the two models.

\section{CONCLUSIONS}

Two different types of models have been presented in this article, a physical spring-block model and various point process models. They are employed a lot in seismology both for the intrinsic description of the physical phenomenon and for the practical use of statistical forecasting. Trying to bridge the two types of models could strengthen both sides. In particular, in this study, a numerical catalog was produced by means of a cellular automata algorithm from a spring-block model based on \cite{PLLR20}. This catalog has been fitted with different types of point process models. On a side, the self correcting model and the stress release model, which take into account that when the fault rupture the amount of strain present around the earthquake location decreases. On the other side, the self-exiting model, in which is implemented the temporal clustering property of the earthquakes. In the middle, there exist a mixing of the two previous type of PP models, called Epidemic Type Aftershock Long-term Correcting models. They take into account both the short term triggering effect and the release of strain by earthquake occurrence. In the triggering model, the background seismicity $\mu$ is assumed constant, whereas for the other models $\mu$ is a function of time. By means of the LogLikelihood maximization procedure, it is possible to find the best set of parameters of each model which makes it better fit the synthetic catalog. \\
The results of this study show that the choice of the best statistical model is not trivial and could depend on the minimum seismic moment at which the seismic catalog is cut. In fact, large events are present in the catalog, it is necessary to introduce the stress release ingredient in order to better fit the experimental catalog. It is important to underline that a numerical catalog is characterized by the absence of incompleteness. Thanks to this, it is possible to attribute this not trivial result certainly not to the incompleteness of the experimental data, but to clustering phenomena and/or preponderant background effects. 
We hope that our results will contribute to develop a new generation PP models.



\begin{table*}
\caption{\label{tab:table3}The parameters of the models used in this article obtained by means of Maximum Likelihood Estimation procedure.}
\begin{ruledtabular}
\begin{tabular}{cccccccc}
 Model & $\alpha$ & $\beta$ & $\xi$ & $\phi$ & c & $\theta $ & $AIC$\\  \hline
 Self-Correcting & $6.45$ & $0.11$ & $6.65 \times 10^{-4}$ & - & - & - & $-24942$ \\
 Stress Release & $6.28$ & $0.15 $ & $1.55 \times 10^{-5}$ & - & - & - & $-26202$\\
 Trigger (TR1)  & $-8.41$  & - & - & $52.25$ & - & $ 150.10 $ & $ -33004 $  \\
 ETAS  & $-9.69$  & - & - & $0.0016$ & $ 6.5 \times 10^{-4} $ & $1.34$ & $-39032 $  \\
 ETASLC1 & $-9.69$ & $-4428$ & $ 0.04 $ & $ 743$ & - & $ 3035 $ & $-41766$ \\
 ETASLC2 & $ -9.69 $ & $ -4129 $ & $ 0.037 $ & $ 1.47 \times 10^{-4}$ & $7.16 \times 10^{-4} $ &  $1.42$ &  $ -44290 $ 
 \label{param}
\end{tabular}
\end{ruledtabular}
\end{table*}

\begin{table*}
\caption{\label{tab:table3}The parameters of the models used in this article obtained by means of Maximum Likelihood Estimation procedure for the numerical catalog with $M \leq 1000$. }
\begin{ruledtabular}
\begin{tabular}{cccccccc}
 Model & $\alpha$ & $\beta$ & $\xi$ & $\phi$ & c & $\theta $ & $AIC$\\  \hline
 Self-Correcting & $6.74$ & $0.18$ & $ 0.001 $ & - & - & - & $ -24920 $ \\
 Stress Release & $6.06$ & $0.15$ & $4.5 \times 10^{-5}$ & - & - & - & $-24896$\\
 Trigger (TR1)  & $-8.20$  & - & - & $ 55.73 $ &  - & $ 157.28 $ & $ -33104 $  \\
 ETAS  & $-9.69$  & - & - & $ 0.002 $ & $3.6 \times 10^{-4}$ & $ 1.26 $ & $ -38874 $  \\
 ETASLC1 & $ -9.69 $ & $ -4478 $ & $ 0.042 $ & $ 1018 $ & - & $ 3755 $ & $ -39104 $ \\
 ETASLC2 & $-9.69$ & $-4129$ & $ 0.038 $ & $ 1.81 \times 10^{-4}$ & $ 4.08 \times 10^{-4} $ & $1.31$ & $-39002$ 
 \label{param2}
\end{tabular}
\end{ruledtabular}
\end{table*}

\begin{table} [htp]
        \centering
        \begin{tabular}{|c|c|c|}
            \hline Parameter & Initial Value & Std. Dev. \\
            \hline $\alpha$ &    0.1     &  0.1      \\
            \hline $\beta$  & 0.1       &  0.1        \\
            \hline $\xi$  & 0.1       &  0.1        \\
            \hline
        \end{tabular}
        
        \caption*{Initial values and standard deviations for the SC model parameters}
        \vspace{5mm}    
 
        \centering
        \begin{tabular}{|c|c|c|}
            \hline Parameter & Initial Value & Std. Dev. \\
            \hline $\alpha$ &    0.1     &  0.1      \\
            \hline $\beta$  & 0.1       &  0.1        \\
            \hline $\xi$  & 0.1       &  0.1        \\
            \hline
        \end{tabular}
     
        \caption*{Initial values and standard deviations for the SR model parameters}
        \vspace{5mm}    
        
       \centering
        \begin{tabular}{|c|c|c|}
            \hline Parameter & Initial Value & Std. Dev. \\
            \hline $\alpha$ &    1.0     &  0.1      \\
            \hline $\phi$  & 1.0   &  0.1        \\
            \hline $\theta$  & 1.0       &  0.1        \\
            \hline
        \end{tabular}
      
        \caption*{Initial values and standard deviations for the TR1 model parameters}
        \vspace{5mm}        

          \centering
        \begin{tabular}{|c|c|c|}
            \hline Parameter & Initial Value & Std. Dev. \\
            \hline $\alpha$ &    1.0     &  0.1      \\
            \hline $\xi$  & 0.1  &  0.1        \\
            \hline $\phi$  & 1.0   &  0.1        \\
            \hline $c$  & 1.0   &  0.1        \\
            \hline $\theta$  & 1.0       &  0.1        \\
            \hline
        \end{tabular}
    
        \caption*{Initial values and standard deviations for the ETAS model parameters}
        \vspace{5mm}       
        \label{tabpar}

        \centering
        \begin{tabular}{|c|c|c|}
            \hline Parameter & Initial Value & Std. Dev. \\
            \hline $\alpha$ &    1.0     &  0.1      \\
            \hline $\beta$ &    1.0     &  0.1      \\
            \hline $\xi$  & 0.1  &  0.1        \\
            \hline $\phi$  & 1.0   &  0.1        \\
            \hline $\theta$  & 1.0       &  0.1        \\
            \hline
        \end{tabular}

        \caption*{Initial values and standard deviations for the STLC1 model parameters}
        \vspace{5mm}       
        
          \centering
        \begin{tabular}{|c|c|c|}
            \hline Parameter & Initial Value & Std. Dev. \\
            \hline $\alpha$ &    1.0     &  0.1      \\
            \hline $\beta$ &    1.0     &  0.1      \\
            \hline $\xi$  & 0.1  &  0.1        \\
            \hline $\phi$  & 1.0   &  0.1        \\
            \hline $c$  & 1.0   &  0.1        \\
            \hline $\theta$  & 1.0       &  0.1        \\
            \hline
        \end{tabular}
       
        \caption*{Initial values and standard deviations for the STLC2 model parameters}
        \vspace{5mm}       
        
  \caption{Initial values employed for the MCMC algorithm in all the different models.}
  \label{tabpar}
\end{table}

\begin{acknowledgments}
This research activity has been supported by MEXT Project for Seismology TowArd Research innovation with Data of Earthquake (STAR-E Project), Grant Number: JPJ010217. E. Lippiello acknowledges support from project PRIN201798CZLJ and from VALERE project of the University of Campania “L. Vanvitelli”.
\end{acknowledgments}

\newpage

\appendix

\section{\label{app1}LogLikelihood Maximization procedure}
For the LL function maximization, we employ a MCMC algorithm. Calling $\Theta^{k}=\{\theta_i^{k}\}$ the parameter set for model $k$, an initial value of $\vec{\theta}_0^{k}$ is chosen. Then, the algorithm is carried out in the subsequent steps. \\

\begin{itemize}
    \item Computation of $LL(Y|\theta_0^{k})$. \\
    \item A parameter $\theta_{i^*}^{k}$ is randomly chosen and updated following the rule  $\theta_{i^*}^{k} \rightarrow \theta_{i^*}^{k} + \epsilon_{i^*}$, where $ \epsilon_{i^*}$ is tuned by hand in order to obtain both a convergence of the procedure and avoiding the Markov chain moving very slow. \\
    \item Compare $LL(Y|\theta_i,...,\theta_{i^*}^{k},...)$ and $LL(Y|\vec{\theta}_0^{k})$. If $LL(Y|\theta_i,...,\theta_{i^*}^{k},...) > LL(Y|\vec{\theta}_0^{k})$, then $LL(Y|\vec{\theta}_0^{k}) = LL(Y|\theta_i,...,\theta_{i^*}^{k},...)$ and repeat the previous point. Otherwise keep the old $LL(Y|\vec{\theta}_0^{k})$ and repeat the previous point. 
\end{itemize}
The procedure stops when it is no longer possible to find a likelihood value greater than the previous one. To ensure the convergence of the algorithm, we produce a Markov chain composed by $100000$ steps.

\section{\label{app2}LogLikelihood estimation}
\subsection{SC and SR model}
For the SC model, the intensity of the non-stationary Poisson point process is Eq.(\ref{SC}). Then, the Likelihood function for the Self-Correcting model can be written as

\begin{align}
    L(Y|\alpha,\beta,\xi)= \left(\prod_{i=1}^N e^{\alpha + \beta t_i - \xi \mathcal{N}[0,t_i)}\right) \times \nonumber \\
    e^{-\int_{t_{in}}^{t_{fin}} \exp(\alpha + \beta t - \xi \mathcal{N}[0,t))dt}
\end{align}

To simplify calculation, the Likelihood function is converted to the Log-Likelihood (LL) function:

\begin{align}
    LL(Y|\alpha,\beta,\xi)= \left(\sum_{i=1}^N \alpha + \beta t_i - \xi \mathcal{N}[0,t_i)\right) \nonumber \\
    -\int_{t_{in}}^{t_{fin}} e^{\alpha + \beta t - \xi \mathcal{N}[0,t)}dt
\end{align}

\begin{figure}[b]
\includegraphics[width=9cm]{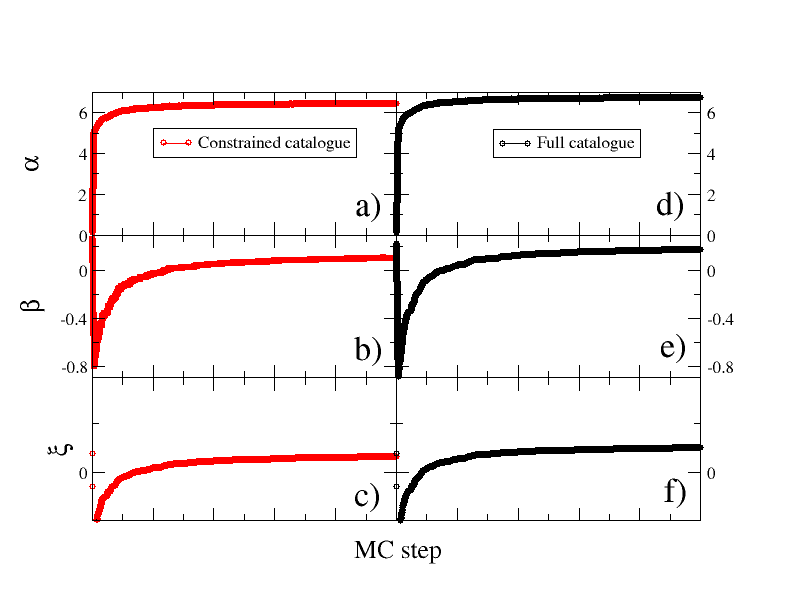}
\caption{\label{fig:epsart} Parameters optimization for the Self-Correcting model. Red curves (a), (b) and (c) represents the evolution of the parameters $\alpha, \beta, \xi$, respectively, during the MCMC algorithm for the catalogue with only $ M \leq 1000 $. Black curves (d), (e) and (f) represents the evolution of the parameter during the MCMC algorithm for the whole numerical catalogue.}
\label{llsc}
\end{figure}

\begin{figure}[b]
\includegraphics[width=9cm]{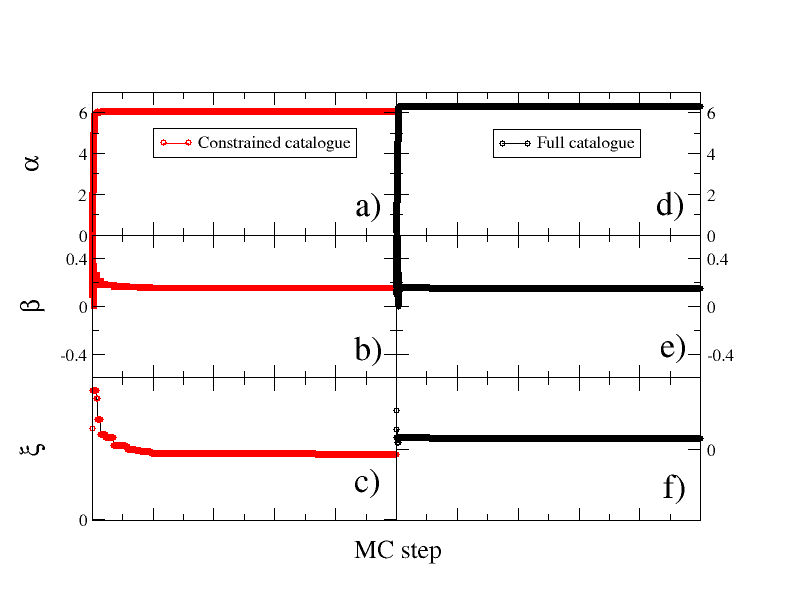}
\caption{\label{fig:epsart} Parameters optimization for the Stress Release model. Red curves (a), (b) and (c) represents the evolution of the parameter during the MCMC algorithm for the catalogue with only $ M \leq 1000 $. Black curves (d), (e) and (f) represents the evolution of the parameter during the MCMC algorithm for the whole numerical catalogue.}
\label{llsr}
\end{figure}

In a very similar way, we can obtain the $ LL(Y|\alpha,\beta,\xi)$ related to the SR model considering $\mathcal{N}[0,t_i) = \mathcal{S}[0,t_i) $ in the definition of $\lambda$. The optimized parameters are listed in Tab.(\ref{param}) whilst the MCMC dynamics for each parameter of the model is plotted in Fig.(\ref{llsc},\ref{llsr}).
     
\subsection{Triggering model}

As for the SC and SR models, the Log-Likelihood function for the triggering model (TR1) \ref{hawkes2},  can be written as 

\begin{align}
    LL(Y|\alpha,\beta,\phi,\theta)= \sum_{i=1}^N \log \left( e^\alpha + \sum_{j:t_j<t_i} \phi e^{-\theta (t_i-t_j)} \right) \nonumber \\
    -\int_{t_{in}}^{t_{fin}} \left( e^\alpha + \sum_{j:t_j<t} \phi e^{-\theta (t-t_j)} \right) dt
    \label{lltreq}
\end{align}

Whereas, for ETAS, we obtain

\begin{align}
    LL(Y|\alpha,\beta,\phi,\theta)= \sum_{i=1}^N \log \left(  e^\alpha + \sum_{j:t_j<t} \frac{\phi}{(t-t_j+c)^\theta} \right) \nonumber \\
    -\int_{t_{in}}^{t_{fin}} \left(  e^\alpha + \sum_{j:t_j<t} \frac{\phi}{(t-t_j+c)^\theta} \right) dt
    \label{lltreq}
\end{align}

The best series $(\alpha,\phi,\theta)$, and $(\alpha,\phi,c,\theta)$ compatible with the recorded data $Y$ are the ones that maximize the likelihood. The optimized parameters are listed in Tab.(\ref{param}). \\

\begin{figure}[b]
\includegraphics[width=9cm]{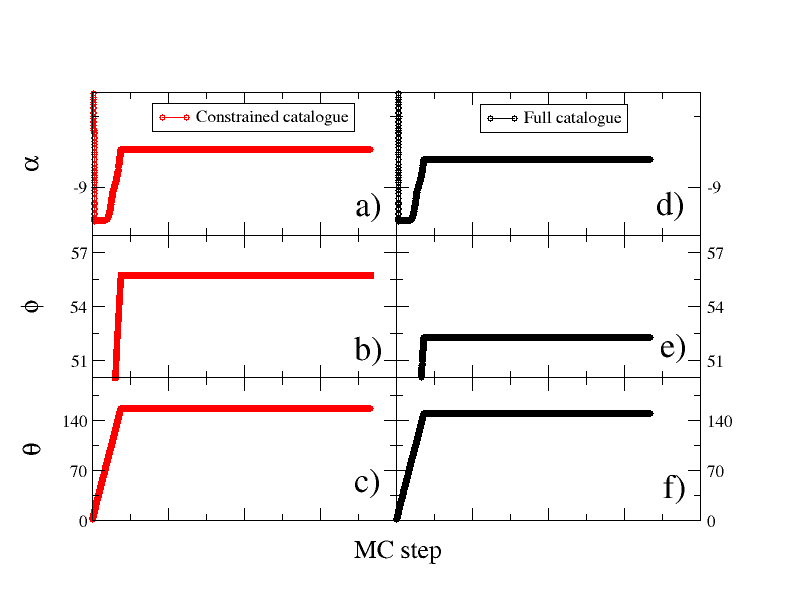}
\caption{\label{fig:epsart} Parameters optimization for the TR1 model. Red curves (a), (b) and (c) represents the evolution of the parameter during the MCMC algorithm for the catalogue with only $ M \leq 1000 $. Black curves (d), (e) and (f) represents the evolution of the parameter during the MCMC algorithm for whole numerical catalogue.}
\label{lltr1}
\end{figure}

\begin{figure}[b]
\includegraphics[width=9cm]{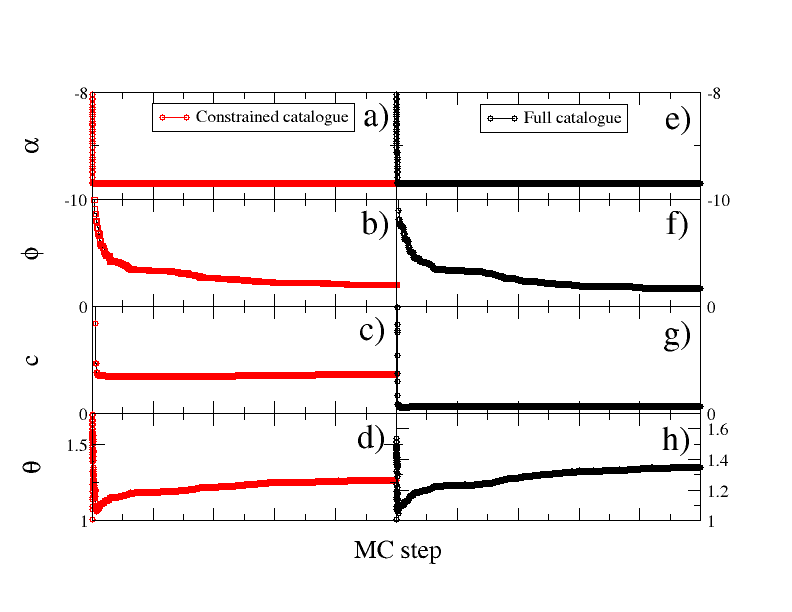}
\caption{\label{fig:epsart} Parameters optimization for the ETAS model. Red curves (a), (b), (c) and (d) represents the evolution of the parameter during the MCMC algorithm for the catalogue with only $ M \leq 1000 $. Black curves (e), (f), (g) and (h) represents the evolution of the parameter during the MCMC algorithm for whole numerical catalogue.}
\label{lltr2}
\end{figure}

\subsection{STLC model}

The analytic expression of the Log-Likelihood function for the ETASLC1 model employing the exponential triggering form can be written as
\begin{align}
    &LL(Y|\alpha,\beta,\phi,\theta)= \nonumber \\
    &\sum_{i=1}^N \log \left[ e^{\alpha + \beta t_i} + \sum_{j:t_j<t_i} \left( \phi e^{-\theta (t_i-t_j)} - \xi \right) \right] \nonumber \\
    &-\int_{t_{in}}^{t_{fin}} \left[ e^{\alpha + \beta t} + \sum_{j:t_j<t} \left( \phi e^{-\theta (t-t_j)} - \xi  \right) \right] dt
    \label{lltreq}
\end{align}

Whereas, implementing the Omori-Utsu functional form we obtain

\begin{align}
    &LL(Y|\alpha,\beta,\phi,c,\theta)= \nonumber \\
    &\sum_{i=1}^N \log \left[ e^{\alpha + \beta t_i} + \sum_{j:t_j<t_i} \left(  \frac{\phi}{(t-t_j+c)^\theta} - \xi \right) \right] \nonumber \\
    &-\int_{t_{in}}^{t_{fin}} \left[ e^{\alpha + \beta t} + \sum_{j:t_j<t} \left( \frac{\phi}{(t-t_j+c)^\theta} - \xi  \right) \right] dt
    \label{lltreq}
\end{align}

The best series $(\alpha,\beta,\phi,c,\theta,\xi)$ compatible with the recorded data $Y$ are the ones that maximize the likelihood. The optimized parameters are listed in Tab.(\ref{param}). 

\begin{figure}[b]
\includegraphics[width=9cm]{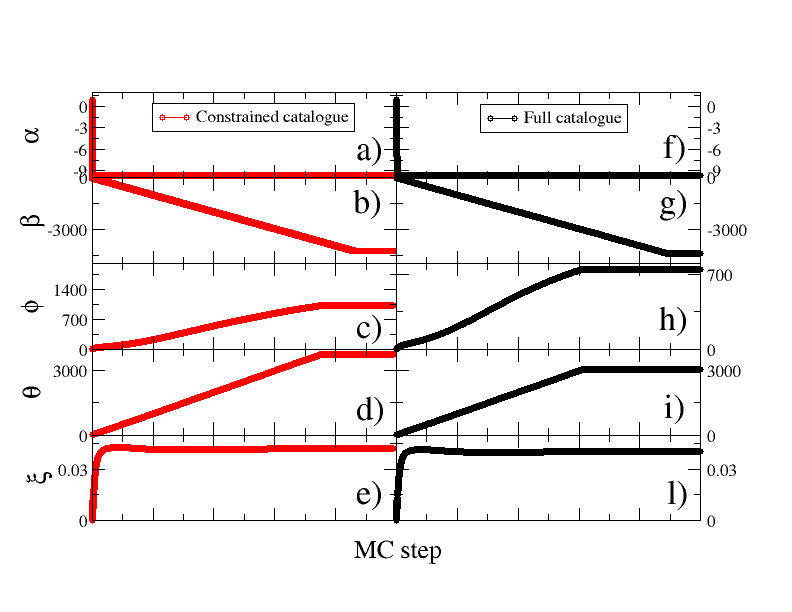}
\caption{\label{fig:epsart} Parameters optimization for the ETASLC1 model. Red curves (a), (b), (c) and (d) represents the evolution of the parameter during the MCMC algorithm for the catalogue with only $ M \leq 1000 $. Black curves (e), (f), (g) and (h) represents the evolution of the parameter during the MCMC algorithm for whole numerical catalogue.}
\label{llselc1}
\end{figure}

\begin{figure}[b]
\includegraphics[width=9cm]{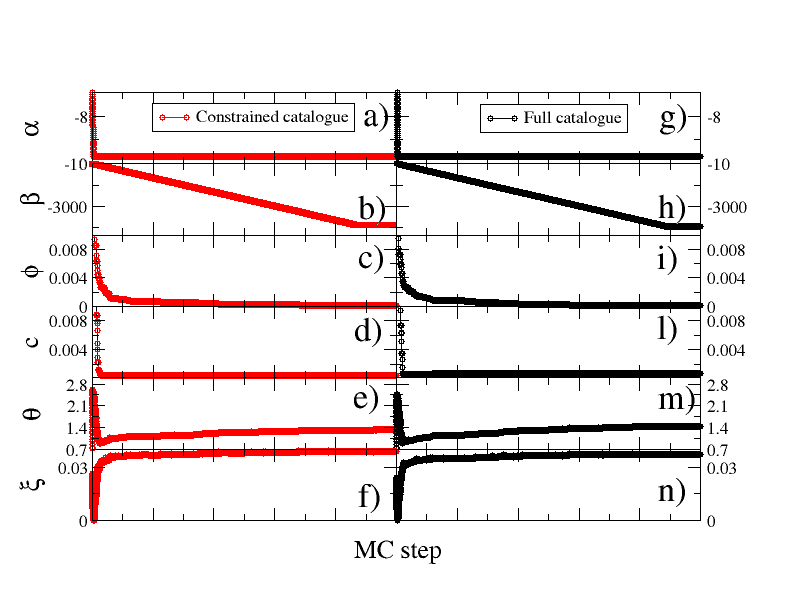}
\caption{\label{fig:epsart} Parameters optimization for the ETASLC2 model. Red curves (a), (b), (c), (d) and (e) represents the evolution of the parameter during the MCMC algorithm for the catalogue with only $ M \leq 1000 $. Black curves (f), (g), (h) and (i) represents the evolution of the parameter during the MCMC algorithm for whole numerical catalogue.}
\label{llselc2}
\end{figure}


\bibliography{apssamp}

\end{document}